\documentclass[aps,prd,twocolumn,groupedaddress,nofootinbib,amssymb,eqsecnum,showpacs,epsfig]{revtex4}
\usepackage{graphicx}
\usepackage{bm}
\usepackage{dcolumn}
\usepackage{amsmath}

\def \lleq {\lower0.9ex\hbox{ $\buildrel < \over \sim$} ~}
\def \ggeq {\lower0.9ex\hbox{ $\buildrel > \over \sim$} ~}

\def \beq  {\begin{equation}}
\def \eeq  {\end{equation}}
\def \ber  {\begin{eqnarray}}
\def \eer  {\end{eqnarray}}

\begin{document}
\newcommand{\newc}{\newcommand}

\newcommand{\ben}{\begin{eqnarray}}
\newcommand{\een}{\end{eqnarray}}
\newc{\be}{\begin{equation}}
\newc{\ee}{\end{equation}}
\newc{\ba}{\begin{eqnarray}}
\newc{\ea}{\end{eqnarray}}
\newc{\bea}{\begin{eqnarray*}}
\newc{\eea}{\end{eqnarray*}}
\newc{\D}{\partial}
\newc{\ie}{{\it i.e.} }
\newc{\eg}{{\it e.g.} }
\newc{\etc}{{\it etc.} }
\newc{\etal}{{\it et al.}}
\newcommand{\nn}{\nonumber}
\newc{\ra}{\rightarrow}
\newc{\lra}{\leftrightarrow}
\newc{\lsim}{\buildrel{<}\over{\sim}}
\newc{\gsim}{\buildrel{>}\over{\sim}}
\title{Newton's constant in $f(R,R_{\mu\nu}R^{\mu\nu},\Box R)$ theories of gravity and constraints from BBN.}
\author{Savvas Nesseris$^1$}
\author{Anupam Mazumdar$^{2,1}$}
%\email{nesseris@nbi.dk}
%\email{anupamm@nbi.dk}
\affiliation{ $^1$
The Niels Bohr International Academy, The Niels Bohr Institute,
Blegdamsvej-17, DK-2100, Copenhagen \O, Denmark\\
$^2$ Physics Department,
Lancaster University, Lancaster, LA1 4YB, UK }
\date{\today}

\begin{abstract}
We consider corrections to the Einstein-Hilbert action which
contain both higher order and nonlocal terms. We derive an
effective Newtonian gravitational constant applicable at the weak
field limit and use the primordial nucleosynthesis (BBN) bound and
the local gravity constraints on $G_{eff}$ in order to test the
viability of several cases of our general Lagrangian. We will also
provide a BBN constrain on the $\Box R$ gravitational correction.
\end{abstract}


\maketitle

\section{Introduction}

One would naturally expect corrections to the Einstein-Hilbert
action of gravity at scales close to the 4-dimensional Planck
scale. However, the details of these corrections in a general time
dependent background are less known. Thus, one would expect a
generic action of type $f(R,
R_{\mu\nu}R^{\mu\nu},R_{\mu\nu\rho\sigma}R^{\mu\nu\rho\sigma},R\Box
R)$~\cite{DeWitt, DeWitt:2007mi,Burgess:2003jk}.

Of these examples, $f(R)$ theories have received much attention
due to their capability to mimic the late-time acceleration,
see~\cite{Nojiri:2008nt,Nojiri:2006ri,Sotiriou:2008ve,Sotiriou:2008rp,Durrer:2007re}
including the solar system constraints~\cite{Nojiri:2003ft}. On
the other hand the nonlocal higher derivative corrections of the
type $R+\sum_{i}R\Box^{i} R$ yield an asymptotically free and
ghost free nonperturbative action of gravity
~\cite{Biswas:2005qr}, which has played an important role in
resolving the big bang singularity in the
Friedmann-Robertson-Walker (FRW) universe. It also explains the
observed temperature anisotropy in a noninflationary bouncing
universe setup~\cite{Biswas:2006bs}. Models of this type were also
studied in \cite{Gottlober:1989ww}, where it was shown that it is
conformally equivalent to Einstein gravity coupled to two scalar
fields. Also, models with nonlocal corrections but with negative
powers of the d'Alambert operator have been considered in
\cite{Nojiri:2007uq}, where it was shown that such a theory may
lead to the unification of early time inflation with late-time
cosmic acceleration.

However, in this paper we will not consider the analysis for
infinite, higher derivative nonlocal corrections of the type
$\sum_{i}R\Box^{i} R$, rather we will only concentrate on the
$i=1$ case, and the $\Box R$ case. The complete nonperturbative
action will be dealt with separately in a future publication.

The aim of the present paper is to study the low scale and long
range behavior of a generic class of Lagrangian density,
$\frac{1}{2} f(R,R_{\mu \nu} R^{\mu\nu},\Box R)$, where we derive
the scalar Newtonian potentials in a homogeneous and an isotropic
expanding background such as in a FRW cosmology. The perturbations
in the FRW background yield a Newtonian potential for a matter
distribution and therefore determine an effective Newtonian
constant $G_{eff}$. At long ranges the linear perturbation
analysis differentiates Einstein's gravity with respect to any
modification through the time evolution of the gravitational
constant, see \cite{Nesseris:2007pa}, \cite{Nesseris:2006er}, and
this is one of the most important differences between Einstein and
modified gravity theories~\footnote{The modifications in general
relativity also affects structure formation~
\cite{Peebles:2004qg,Stabenau:2006td,Koivisto:2006ie} and the
predictions in the cosmic microwave background radiation through
radiation-matter equality~ \cite{Liddle:1998ij,Chen:1999qh}. We
will study various consequences to structure formation and Cosmic
microwave background radiation in a separate publication.}.

This difference can be tested by using the primordial
nucleosynthesis (BBN) bounds on the gravitational constant, which
are of the order of $10\%$
 \cite{Bambi:2005fi,Iocco:2008va,Clifton:2005xr,Coc:2006rt}. The
BBN bounds are important due to the fact that the value of the
gravitational constant determines the expansion rate of the
Universe and thus the relevant time scales for the production of
light elements (H, He and Li), see~\cite{Coc:2006rt}. As a
consequence, if we assume that the gravitational constant at the
time of BBN is different from its value today, this means that the
light element abundances will be different with respect to the
standard BBN predictions. Even a weak time dependence, which gives
no observable effects in Solar system experiments performed at the
present epoch and at small scales, could give observable effects
when translated over cosmological time scales. So, it will be
interesting to analyze some special cases of our general
Lagrangian and use the BBN bounds on the gravitational constant to
place constraints on the parameters of these simple models.

Furthermore, we will be applying the BBN constraints to study the
$\Box R$ corrections in the Einstein-Hilbert action. Previous
studies of nonlocal action has concentrated on formal aspects of
the validity of effective field theory~\cite{Barvinsky:2002uf} and
particle creation~\cite{Dobado:1998mr}. It should be noted that
the effective Newton's constant in nonlocal gravity and its
implications to cosmology, BBN and the Solar System have been also
considered in \cite{Wetterich:1997bz} and more recently in models
generalizing this in \cite{Koivisto:2008dh}.

Here we consider the alteration of classical dynamics of the
Universe due to the presence of the $\Box R$ gravitational
correction.

%%%%%%%%%%%%%%%%%%%%%%%%%%%%%%%%
\section{Background equations}

The action we will consider is
\be S=\int d^4x
\sqrt{-g}\left[\frac{1}{2} f(R,R_{\mu \nu} R^{\mu \nu},\Box
R)+L_m\right] \label{action}
\ee
where $R$ is the Ricci scalar,
$R_{\mu \nu}$ the Ricci tensor, $\Box$ is the d'Alembert operator
$\Box \equiv g^{\alpha\beta} \nabla_\alpha \nabla_\beta$ and $L_m$
the matter Lagrangian. We use the metric signature $(-,+,+,+)$.

Varying the action with respect to the metric $g^{\mu \nu}$ we
obtain the field equations as
\cite{Carroll:2004de},\cite{Schmidt:1990dh}:
\ba \hspace{-.5cm}&& \left(F+\Box \frac{\partial f}{\partial \Box
R}\right) R_{\mu \nu} -\frac{1}{2}g_{\mu \nu} f+ 2 f_{,RR}
g^{\lambda \kappa}R_{\nu \kappa}R_{\mu \lambda}\nn
\\ \hspace{-.5cm}&& -(\nabla _{(\mu} R) ~ \left(\nabla_{\nu)} \frac{\partial f}{\partial \Box
R}\right)+\frac{1}{2}g_{\mu \nu}\left[R^{;\sigma}\frac{\partial
f}{\partial \Box R}\right]_{;\sigma}+ \nn \\\hspace{-.5cm}
&&\left[g_{\mu \nu} \Box-\nabla_\mu \nabla_\nu\right]\left(F+\Box
\frac{\partial f}{\partial \Box R}\right)+\Box \left(f_{,RR}
R_{\mu \nu}\right)+ \nn
\\\hspace{-.5cm} && g_{\mu \nu}\nabla_\alpha\nabla_\beta
\left(f_{,RR} R^{\alpha \beta}\right)-2\nabla_\alpha\nabla_\beta
\left(f_{,RR}~R^\alpha_{~(\mu}~\delta^\beta_{~\nu)}\right) \nn
\\\hspace{-.5cm}&&=T_{\mu \nu} \label{fieldeqs} \ea
where $F=\frac{\partial f}{\partial R}$ and
$f_{,RR}=\frac{\partial f}{\partial (R_{\mu \nu} R^{\mu \nu})}$.
We have also defined the energy-momentum tensor as \be
T_{\mu\nu}=-\frac{2}{\sqrt{-g}}\frac{\delta(\sqrt{-g} L_m)}{\delta
g^{\mu\nu}} \label{en-mom-tensor}\ee and the parentheses next to
indices mean symmetrization, e.g.
$A_{(ij)}=\frac{1}{2}\left(A_{ij}+A_{ji}\right)$. Note that by
using the field Eqs. (\ref{fieldeqs}) it is easy to see that it
has the correct limits, i.e. in the case when the Lagrangian is
given by $f=R+\xi \Box R$ then we get general relativity, as the
$\Box R$ term can be written as a total divergence. Also, for a
conformally flat metric and a Lagrangian given
by\cite{Barrow:1983rx} $f=R+\xi (3R_{\mu \nu}R^{\mu \nu}-R^2)$ the
field equations give general relativity at the background level,
but not at the perturbations level as the metric is no longer
conformally flat.

In a flat FRW metric with a scale factor $a(t)$, we obtain the
zero-order (background) equations:
\begin{widetext}
\ba && \frac{f}{2}-3 F \left(H^2+\dot{H}\right)+3 \dot{F} H -9
H^3\left( 2f_{,RR} H-2 \dot{f}_{,RR}- \dot{\frac{\partial
f}{\partial \Box R}} \right)+36 \dot{H} H^2\left( \frac{\partial
f}{\partial \Box R} +\frac{1}{2}f_{,RR} \right) -6
\ddot{\frac{\partial f}{\partial \Box R}}
H^2+\nn \\
&& 3 H \ddot{H} \left(7\frac{\partial f}{\partial \Box R}+ 4
f_{,RR} \right)+12  \dot{H} H \left(\dot{f}_{,RR} -
\dot{\frac{\partial f}{\partial \Box R}} \right)-3
\frac{d^3\frac{\partial f}{\partial \Box R}}{dt^3}
H+12\dot{H}^2\left(\frac{\partial f}{\partial \Box R} -
f_{,RR}\right)-3 \ddot{H} \dot{\frac{\partial f}{\partial \Box R}}+\nn \\
&&3 \dot{H} \ddot{\frac{\partial f}{\partial \Box R}} +3
\frac{\partial f}{\partial \Box R} \frac{d^3 H}{dt^3} =\rho
\label{friedeq1}\ea
\ba && -2 F \dot{H}-\ddot{F}-3 \left(2
\ddot{f}_{,RR}+\ddot{\frac{\partial f}{\partial \Box R}}\right)
H^2+\left(\dot{F}-12 \ddot{H} f_{,RR}-24 \dot{H} \dot{f}_{,RR}-21
\dot{H} \dot{\frac{\partial f}{\partial \Box R}}+2
\frac{d^3\frac{\partial f}{\partial \Box R}}{dt^3}\right) H \nn \\
&&+6 \dot{f}_{,{RR}} H^3 -4 \ddot{f}_{,{RR}} \dot{H}-
\ddot{H}\left(8 \dot{f}_{,RR}+3\dot{\frac{\partial f}{\partial
\Box R}}\right)-24 \dot{H}^2 f_{,{RR}}+8\dot{H}
\ddot{\frac{\partial f}{\partial \Box R}}-4 f_{,{RR}}
\frac{d^3H}{dt^3}+\frac{d^4\frac{\partial f}{\partial \Box
R}}{dt^4}=\rho \ea
\end{widetext}
where the dot $(~\dot{}~)$ denotes a derivative with respect to
time, eg. $\dot{f}_{,RR}\equiv \frac{\partial f_{,RR}}{\partial
t}$ and $f_{,RR}=\frac{\partial f}{\partial (R_{\mu \nu} R^{\mu
\nu})}$.

\section{Perturbation equations}

We will consider the following perturbed metric with scalar metric
perturbations $\Phi$ and $\Psi$ in a longitudinal gauge:
\be
ds^2=-(1+2\Phi)dt^2+a(t)^2(1-2\Psi)\delta_{ij}dx^idx^j
\label{metric}
\ee
The energy-momentum tensor of the nonrelativistic matter is
decomposed as $T_0^0=-(\rho_m+\delta \rho_m)$ and
$T^0_\alpha=-\rho_m \upsilon_{m,\alpha}$, where $\upsilon_{m}$ is
a velocity potential. The Fourier transformed perturbation
equations for the continuity equations are given by
\ba -\frac{\rho _m \upsilon _m k^2}{a}-\dot{\delta \rho _m}-3 H
\delta \rho _m+3 \dot{\Psi } \rho _m =0\label{cont0}
\ea
\ba \Phi  \rho _m-a \left(H \rho _m \upsilon _m+\rho _m
\dot{u}_m\right)=0 \label{cont1} \ea
Following the approach of Refs.
\cite{Tsujikawa:2007gd},\cite{Nesseris:2008mq}, we use a
subhorizon approximation under which the leading terms correspond
to those containing $k^2$ and $\delta \rho_m$. Terms that are of
the form $H^2 \Phi$ or $\ddot{\Phi}$ are considered negligible
relative to terms like $(k^2/a^2)\Phi$ for modes well inside the
Hubble radius $(k^2\gg a^2 H^2)$. Under this approximation, the
Fourier transformed perturbation equations, coming from the
$(\mu,\nu)=(0,0)$ and $(1,2)$ terms of the field Eqs.
(\ref{fieldeqs}), are given by
\ba -\delta \rho _m  &-& \frac{k^4}{a^4} \left({\delta
{\frac{\partial f}{\partial \Box R}}}- 2
f_{,{RR}} (\Phi -\Psi )\right)\nn \\
&+&\frac{k^2}{a^2} ({\delta F}-2 F \Psi )=0 \label{per00} \ea
\ba &&\frac{\delta F}{F}+ \Phi - \Psi+\nn \\
&&~~~~~\frac{k^2}{a^2} \frac{1}{F}\left(- {\delta{\frac{\partial
f}{\partial \Box R}}}+ (\Phi -3 \Psi ) f_{,{RR}}\right)
=0~~~~~~~~~~~~~~\label{per12} \ea

While in general relativity in the case of a matter fluid with no
anisotropic stress the two potentials $\Phi$ and $\Psi$ are equal,
as can be seen from Eq. (\ref{per12}), this is not the case for
modified gravity theories as the gravity sector alone induces an
anisotropic stress and creates the inequality of $\Phi$ and
$\Psi$, see for example \cite{Tsujikawa:2007gd,Nesseris:2008mq}.
Next, we define the gauge invariant matter density perturbation
$\delta_m$ as
 \be \delta_m\equiv \frac{\delta \rho_m}{\rho_m}+3 H
\upsilon \label{ddm}\ee where \be \upsilon= a \upsilon_m \ee
Under this approximation Eqs. (\ref{cont0}),(\ref{cont1}) and
(\ref{ddm}) yield
\ba \ddot{\delta}_m+2 H \dot{\delta}_m
+\frac{k^2\Phi}{a^2}\simeq0 \label{perturb1} \ea
Next, we write $\delta F$ and $\delta \frac{\partial f}{\partial
\Box R}$ as \be \delta F= \frac{\D F}{\D R} \delta R +\frac{\D
F}{\D(R_{\mu\nu}R^{\mu\nu})}\delta (R_{\mu\nu}R^{\mu\nu})+\frac{\D
F}{\D (\Box R)}\delta(\Box R)\ee

\ba \delta \frac{\partial f}{\partial \Box R}&=&\frac{\D F}{\D
\Box R} \delta R+\frac{\D \frac{\partial f}{\partial \Box
R}}{\D(R_{\mu\nu}R^{\mu\nu})}\delta (R_{\mu\nu}R^{\mu\nu})\nn \\
&+&\frac{\D^2 f}{\D (\Box R)^2}\delta(\Box R)~~~~~~~~~~~~~~~~~~~~
\ea where $\delta R$, under the subhorizon approximation, is given
by \be \delta R \simeq -2\frac{k^2}{a^2}\left(2 \Psi-\Phi\right)
\label{dR}\ee while $\delta (R_{\mu\nu}R^{\mu\nu})\sim 0$ and
$\delta(\Box R)$ is given by \be \delta(\Box R)= -\frac{2
k^4}{a^4}\left(\Phi- 2 \Psi \right) \ee

Making these substitutions and using the subhorizon approximation
in Eqs. (\ref{per00}) and (\ref{per12}) we get \ba && -\delta \rho
_m -2 F \frac{k^2}{a^2} \Psi +2 {\frac{\D^2
f}{\partial (\Box R)^2}} \frac{k^8}{a^8} (\Phi -2 \Psi )\nn \\
&&+2 \frac{k^4}{a^4} \left(F_{{,R}}(\Phi -2 \Psi )+ f_{,{RR}}(\Phi
- \Psi )\right)-4 {\frac{\partial F}{\partial \Box R}}
\frac{k^6}{a^6} (\Phi -2 \Psi )\nn \\ && =0\label{eqpp1}\ea

\ba && F (\Phi -\Psi )+ \frac{k^2}{a^2} (\Phi -3 \Psi )
f_{,{RR}}-2 (\Phi -2 \Psi )\cdot\nn \\ &&\left(-\frac{k^2}{a^2}
F_{{,R}}+ 2\frac{\partial F}{\partial \Box R} \frac{k^4}{a^4} -
\frac{\D^2 f}{\partial (\Box R)^2} \frac{k^6}{a^6} \right) =0
\label{eqpp2}\ea

The next step is to express $\Phi$ and $\Psi$ in terms of
$\delta_m$. This can be done by solving the system of Eqs.
(\ref{eqpp1}) and ({\ref{eqpp2}) for $\Phi$ and $\Psi$. Doing so
we find \begin{widetext} \ba &&\Phi = -\frac{a^2}{k^2}\frac{\rho
\delta _m}{2(F-\frac{k^2}{a^2}f_{,{RR}})}\frac{F+ \frac{k^2}{a^2}
\left(3f_{,{RR}}+4F_{{,R}}\right) -8 {\frac{\partial F}{\partial
\Box R}} \frac{k^4}{a^4}+4 {\frac{\D^2 f}{\partial (\Box R)^2}}
\frac{k^6}{a^6} }{F+\frac{k^2}{a^2} \left(
2f_{,{RR}}+3F_{{,R}}\right)-6 {\frac{\partial F}{\partial \Box R}}
\frac{k^4}{a^4}+3 {\frac{\D^2 f}{\partial (\Box R)^2}}
\frac{k^6}{a^6}}\label{phihor} \ea

\ba && \Psi= -\frac{a^2}{k^2}\frac{\rho \delta
_m}{2(F-\frac{k^2}{a^2}f_{,{RR}})}\frac{F+ \frac{k^2}{a^2}
\left(f_{,{RR}}+2F_{{,R}}\right) -4 {\frac{\partial F}{\partial
\Box R}} \frac{k^4}{a^4}+2 {\frac{\D^2 f}{\partial (\Box R)^2}}
\frac{k^6}{a^6} }{F+\frac{k^2}{a^2} \left(
2f_{,{RR}}+3F_{{,R}}\right)-6 {\frac{\partial F}{\partial \Box R}}
\frac{k^4}{a^4}+3 {\frac{\D^2 f}{\partial (\Box R)^2}}
\frac{k^6}{a^6}} \label{psihor}\ea

\end{widetext}

From Eq. (\ref{phihor}) we can define a Poisson equation in the
Fourier space and attribute the extra terms that appear on the
right-hand side to an effective gravitational constant $G_{eff}$.
Doing so, we get the gravitational potential \ba &&\Phi = -4 \pi
G_{eff} \frac{a^2}{k^2} \delta _m \rho _m \label{poisson} \ea
where $G_{eff}$ is defined as
\begin{widetext}
\ba G_{eff}\equiv\frac{1}{8
\pi}\frac{1}{F-\frac{k^2}{a^2}f_{,{RR}}}\cdot\frac{F+
\frac{k^2}{a^2} \left(3f_{,{RR}}+4F_{{,R}}\right) -8
{\frac{\partial F}{\partial \Box R}} \frac{k^4}{a^4}+4 {\frac{\D^2
f}{\partial (\Box R)^2}} \frac{k^6}{a^6} }{F+\frac{k^2}{a^2}
\left( 2f_{,{RR}}+3F_{{,R}}\right)-6 {\frac{\partial F}{\partial
\Box R}} \frac{k^4}{a^4}+3 {\frac{\D^2 f}{\partial (\Box R)^2}}
\frac{k^6}{a^6}}\label{geff} \ea
\end{widetext}
Note that the inclusion of the term $R\Box R$ has a negative
contribution to $G_{eff}$. For certain choice of parameters it
might be possible to make $G_{eff}$ vanishingly small, thereby
modifying the Newtonian gravity on large temporal and spatial
scales.

Since the corrections from different forms of the modifications,
i.e. terms like $R^2$, $R_{\mu\nu} R^{\mu\nu}$, $\Box R$ etc enter
with different powers of the $k^2$ it is interesting to check
which hierarchies exist between the various coefficients in order
for them to be equally important at some interesting scales. This
can be very helpful to understand the relative importance of the
various modifications at different regimes. However, this is
possible only for some simple cases and when the Lagrangian $f$ is
completely specified. In the general case it is not easy to tell
whether a term of an arbitrary function, for example of $R\Box R$
is more important than some other term, as any of the derivatives
of $f$, ie $F$, $F_{,R}$ etc may contain terms like $\Box R$.

On the other hand by studying some simple cases, like the ones
mentioned in the Examples section, we can draw some interesting
conclusions. For example, as can be seen from Eq. (\ref{geff}) for
very small or very large scales $\frac{k}{a}$ the terms containing
$\Box R$ are not as important as terms involving $F$ and
$f_{,RR}$. However, on intermediate scales the $\Box R$ terms can
affect the behavior of $G_{eff}$ and actually enters with a
negative sign which means that it may drive $G_{eff}$ to zero or
an unphysical singularity.

Let us now study the Eq. (\ref{perturb1}) of matter perturbations
\ba &&
\ddot{\delta}_m+2 H \dot{\delta}_m -4 \pi
G_{eff}\rho_m\delta_m\simeq0 \label{perturb2} \ea
Note that the above expression will modify the large scale
structure behavior on small scales as well as on large scales
through higher order modifications. We will study these
interesting possibilities in future publications.

\section{Examples}

In this section, we will consider several examples for the very
general Lagrangian of the action (\ref{action}) in order to
demonstrate how our results can be applied to a vast group of
possible theories.

\subsection{$f(R)$ gravity}

\begin{figure*}[!t]
\centering
\vspace{0cm}\rotatebox{0}{\vspace{0cm}\hspace{0cm}\resizebox{.48\textwidth}{!}{\includegraphics{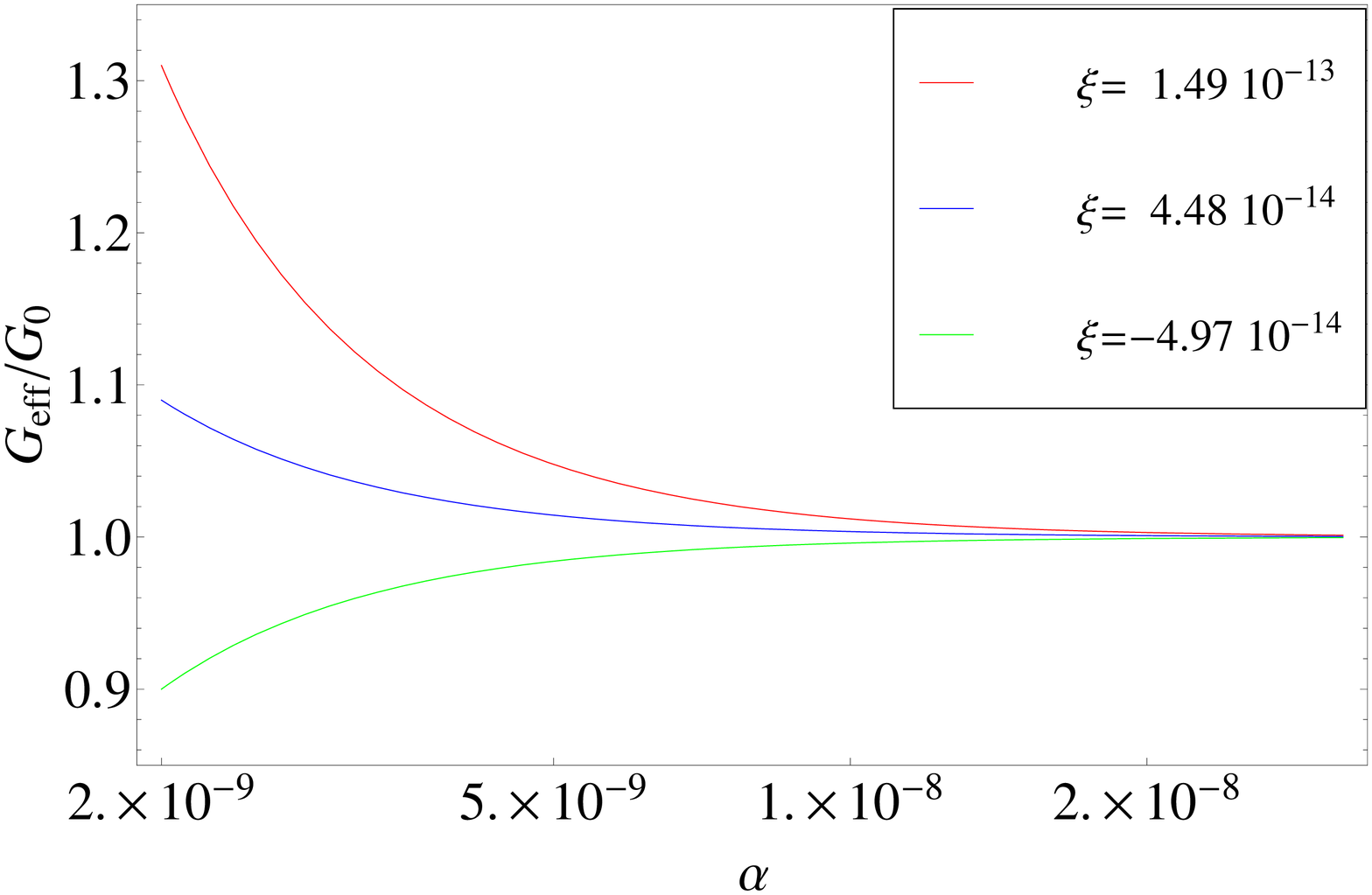}}}
\vspace{0cm}\rotatebox{0}{\vspace{0cm}\hspace{0cm}\resizebox{.48\textwidth}{!}{\includegraphics{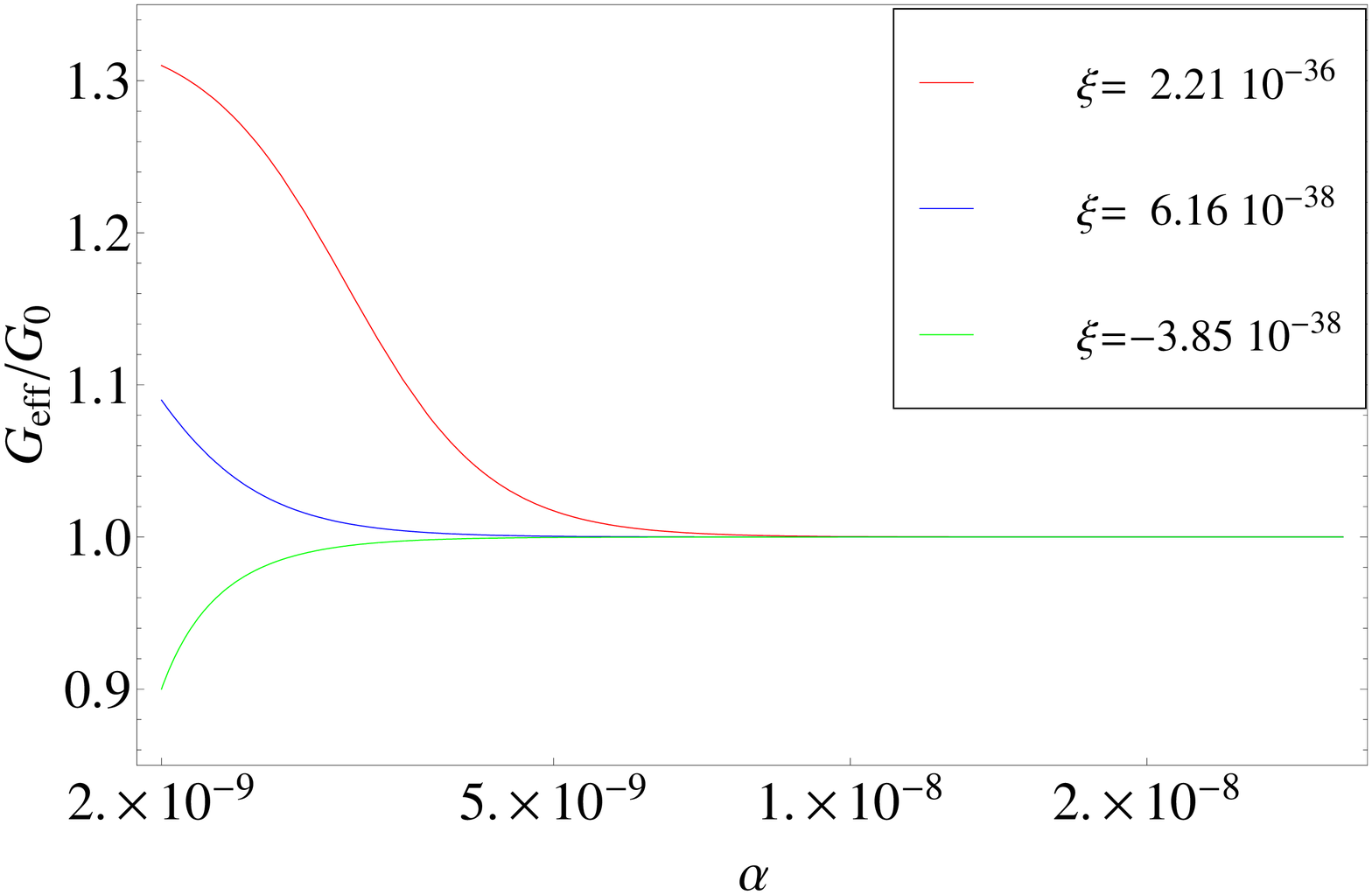}}}
\caption{Plots of $G_{eff}$ as a function of the scale factor $a$,
given by Eq. (\ref{geffex11}) for example B (left) and by Eq.
(\ref{geffex2}) for example C (right). The values of the parameter
$\xi$ (in units of $Mpc^2$ and $Mpc^6$ respectively) used
correspond to the best (blue line) and $1\sigma$ values (green and
red lines) allowed by the BBN bounds and are shown in each legend
respectively. \label{plot1}}
\end{figure*}

As a first example we will consider $f(R)$ theories, for which we
have to set $f(R,R_{\mu \nu} R^{\mu \nu},\Box R)=f(R)$. Then Eq.
(\ref{geff}) yields \be G_{eff}=\frac{1}{8\pi F}
\frac{1+4\frac{k^2}{a^2R}m}{1+3\frac{k^2}{a^2R}m}
\label{GefffR}\ee where \be m\equiv\frac{RF_{,R}}{F}\nn \ee being
in agreement with the standard results from $f(R)$ gravity
\cite{Tsujikawa:2007gd}.

\subsection{$f(R,R_{\mu\nu}R^{\mu\nu})$ gravity }

A second example will be the Lagrangian \be f(R,R_{\mu \nu} R^{\mu
\nu},\Box R)= R+\sum_{n=0}^\infty \xi_n (R_{\mu\nu}R^{\mu\nu})^n
\ee In this case Eq. (\ref{geff}) gives \ba\hspace{-1cm}
G_{eff}(a)=&&\frac{1}{8 \pi}\cdot\frac{1}{1-
\frac{k^2}{a^2}\sum_{n=0}^\infty n \xi_n
(R_{\mu\nu}R^{\mu\nu})^{n-1}}\cdot\nn\\&&\frac{1+3
\frac{k^2}{a^2}\sum_{n=0}^\infty n \xi_n
(R_{\mu\nu}R^{\mu\nu})^{n-1}}{1+2 \frac{k^2}{a^2}\sum_{n=0}^\infty
n \xi_n (R_{\mu\nu}R^{\mu\nu})^{n-1}}\label{geffex1}\ea If we keep
only the first order term of the sum, corresponding to the
Lagrangian $R+\xi R_{\mu\nu}R^{\mu\nu}$, where $\xi$ is a
constant, then $G_{eff}$ is \be G_{eff}(a)=\frac{1}{8 \pi}\cdot
\frac{1+3 \frac{k^2}{a^2}\xi}{(1- \frac{k^2}{a^2}\xi)(1+2
\frac{k^2}{a^2}\xi)}\label{geffex11}\ee

Next we will use the BBN constraints on the variation of the
gravitational constant to constrain the parameter $\xi$. The
effect of the variation of $G_{eff}$ can be constrained from BBN
to be of the order of $10\%$, see, for example, Ref.
\cite{Bambi:2005fi}, which gives
$\frac{G_{BBN}}{G_0}=1.09\pm^{0.22}_{0.19}$. It is possible to use
Eq. (\ref{geffex11}) to find analytically the best and the
$1\sigma$ values of $\xi$ according to BBN \be
\xi_{BBN}=\frac{a_{BBN}^2 \left(-3+\frac{G_{BBN}}{G_0}\pm
\sqrt{9-14 \frac{G_{BBN}}{G_0}+9
(\frac{G_{BBN}}{G_0})^2}\right)}{4 k^2 \frac{G_{BBN}}{G_0}}
\label{xi1}\ee In Fig. 1 (left) we show the plot of $G_{eff}$,
given by Eq. (\ref{geffex1}), for the values of the parameter
$\xi$, which correspond to the central and $1\sigma$ values
allowed by the BBN bounds for a value of $k=0.002 Mpc^{-1}$.
However, since the $k$-mode is actually unknown and can only be
rather arbitrarily chosen, we have also plotted the value of $\xi$
versus $k$ in Fig. 2 and in Table 1 we show $\xi$ for various
values of the scale $k$. It is interesting to note that in this
case there are actually two allowed values of $\xi$ by the BBN
constraints, however only one is shown in Table 1 for each $k$ as
the other results in completely unphysical behavior for $G_{eff}$.
Finally, we also consider the $k$-mode corresponding to the
horizon size at the BBN as the relevant scale. Since the horizon
at the BBN is approximately $\sim 10^{-4} h^{-1} Mpc$, see, for
example, Ref. \cite{Bamba:2008hr}, this corresponds to a scale
$k_{BBN}\sim \frac{a_{BBN}}{\lambda_{hor}} \sim 10^{-5} Mpc^{-1}$
and the corresponding constraints are shown in Table 1.

\begin{figure*}[!t]
\centering
\vspace{0cm}\rotatebox{0}{\vspace{0cm}\hspace{0cm}\resizebox{.48\textwidth}{!}{\includegraphics{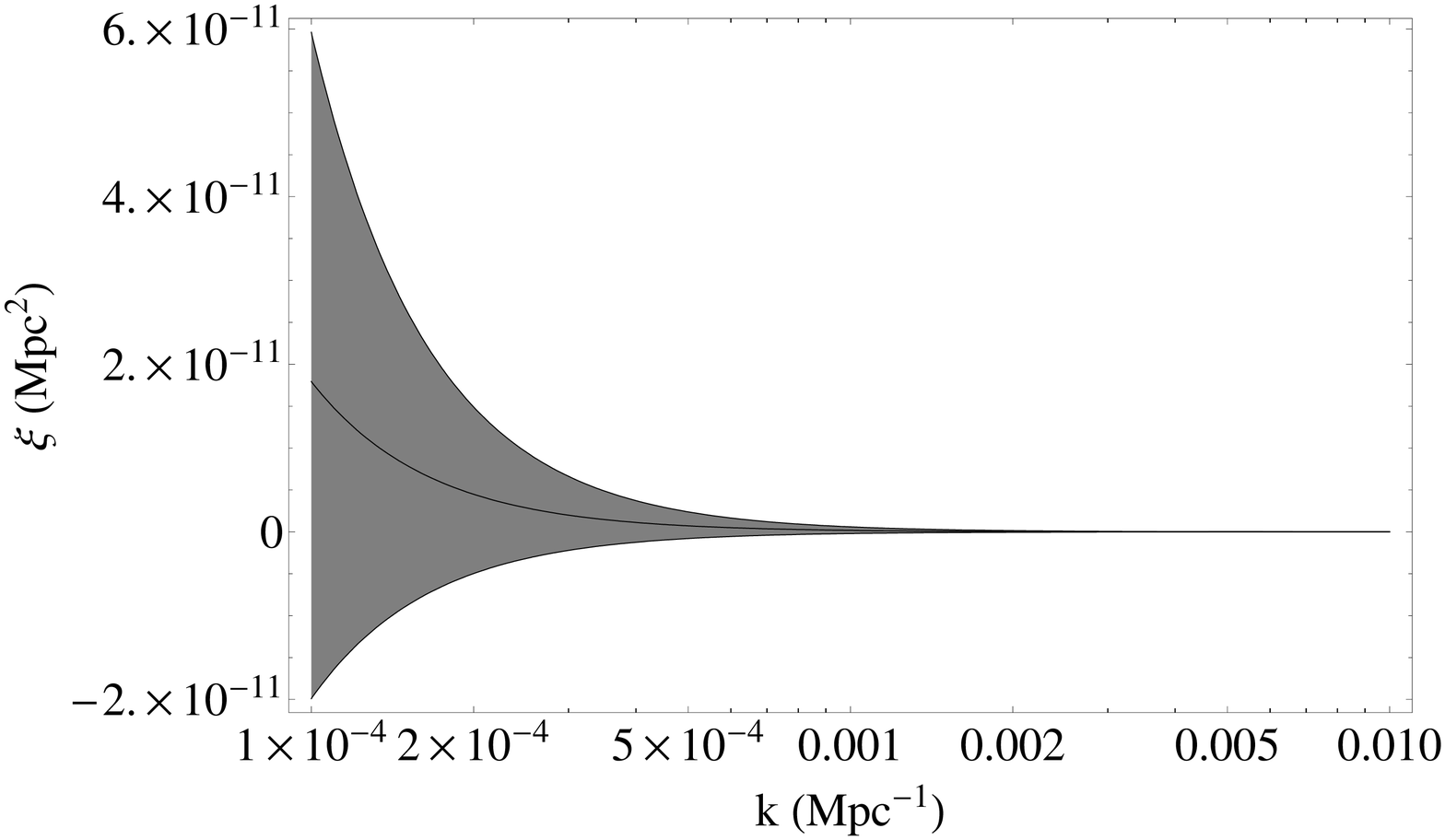}}}
\vspace{0cm}\rotatebox{0}{\vspace{0cm}\hspace{0cm}\resizebox{.48\textwidth}{!}{\includegraphics{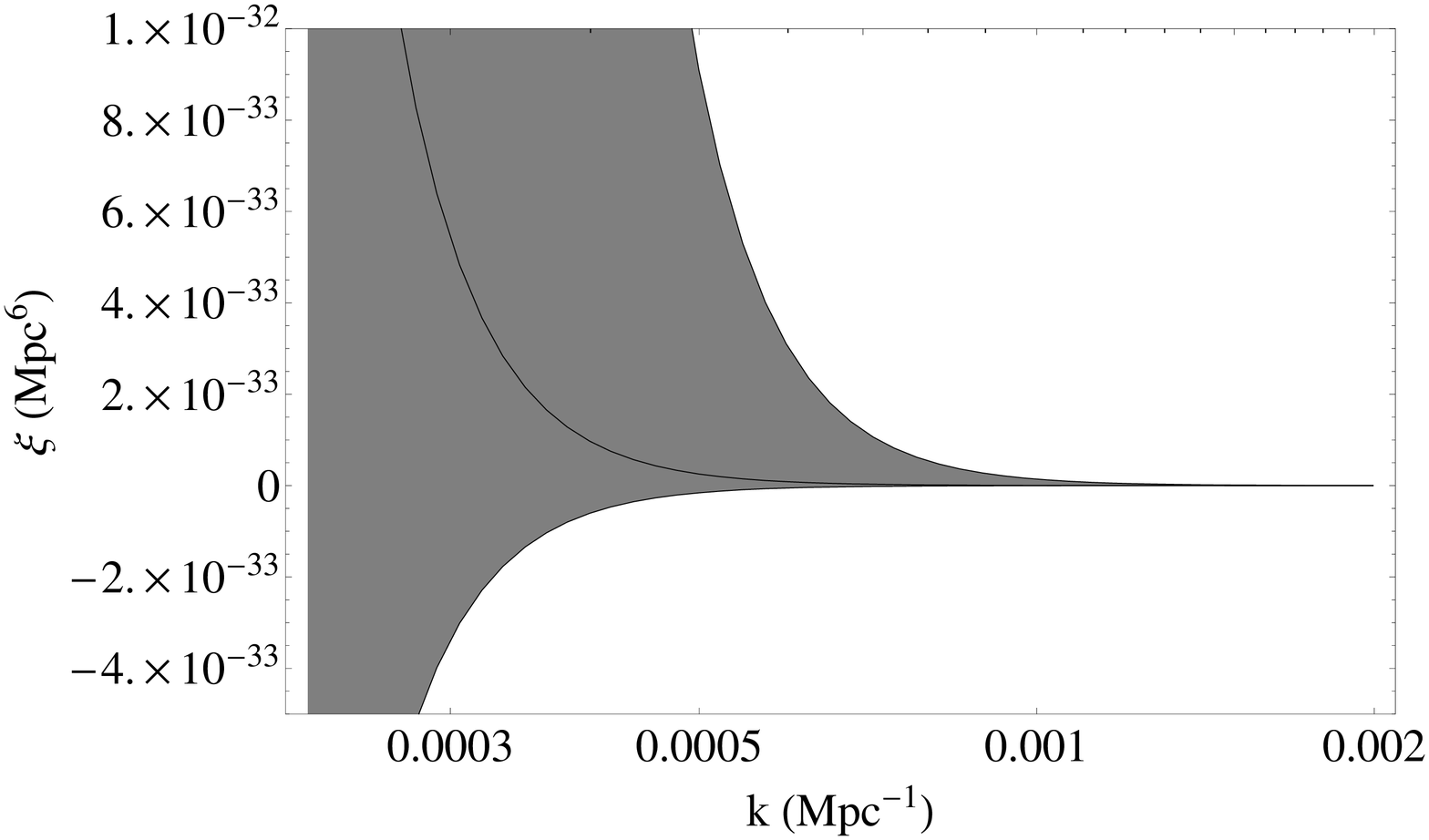}}}
\caption{Plots of $\xi$ as a function of the scale $k$, given by
Eq. (\ref{xi1}) for example B (left) and by Eq. (\ref{xi2}) for
example C (right). The grey areas correspond to the $1\sigma$
error bars. The scale that corresponds to the horizon at the BBN
is $k_{BBN}\sim 10^{-5} Mpc^{-1}$ and is situated outside of the
range of the plots. \label{plot1}}
\end{figure*}

It is possible to get more robust bounds on our models by
considering local gravity constraints following the approach of
Ref. \cite{Tsujikawa:2007gd}. In this case we demand that strong
modifications of gravity should not be observed on scales up to
$\lambda_k \sim a/k$, where in solar system experiments the scale
$\lambda_k$ corresponds to a value around $\lambda_k=1 AU$.
Therefore, taking into consideration Eq. (\ref{geffex11}), we
demand that $\frac{k^2}{a^2} |\xi| \ll 1$, which gives the
following constraint \be |\xi| \ll \lambda_k^2 \sim 10^{-23} Mpc^2
\ee While this is more robust than the ones found by using the BBN
constraint, the latter are not excluded as $\xi_{BBN}$ has a
larger $1\sigma$ error region, so the two constraints overlap with
each other.

\begin{table}[!b]
\begin{center}
\caption{The parameter $\xi$ using the BBN constrain for various
values of the scale $k$. The first entry corresponds to the scale
of the solar system experiments $\lambda_k\sim 1AU$ or
$k_{sol}\sim 2~10^{11} Mpc^{-1}$, while the last ($k\sim
10^{-5}Mpc^{-1}$) corresponds to the scale of the horizon during
the BBN.}
\begin{tabular}{ccc}
\hline
\hline\\
\vspace{6pt}   $k ~(Mpc^{-1})$ &\vspace{6pt}  $\xi ~(Mpc^{2})$ ~(case B)                          & \vspace{6pt}  $\xi ~(Mpc^{6})$ ~(case C) \\
\hline
\hline\\

\vspace{6pt}   $2\cdot 10^{11}$     &\vspace{6pt}  $|\xi| \ll 10^{-23}$                                & \vspace{6pt}  $|\xi| \ll 10^{-68}$ \\

\vspace{6pt}   $1\cdot 10^{-1}$       &\vspace{6pt}  $1.79*10^{-17}\pm ^{4.17*10^{-17}}_{3.78*10^{-17}}$ & \vspace{6pt}  $3.95*10^{-48}\pm ^{1.38*10^{-46}}_{6.41*10^{-48}}$ \\

\vspace{6pt}   $2\cdot 10^{-3}$     &\vspace{6pt}  $4.48*10^{-14}\pm ^{1.04*10^{-13}}_{9.46*10^{-14}}$ & \vspace{6pt}  $6.16*10^{-38}\pm ^{2.15*10^{-36}}_{1.00*10^{-37}}$ \\

\vspace{6pt}   $3\cdot 10^{-4}$     &\vspace{6pt}  $1.99*10^{-12}\pm ^{4.63*10^{-12}}_{4.20*10^{-12}}$ & \vspace{6pt}  $5.41*10^{-33}\pm ^{1.89*10^{-31}}_{8.79*10^{-33}}$ \\

\vspace{6pt}   $1\cdot 10^{-5}$       &\vspace{6pt}  $1.79*10^{-9}\pm ^{4.17*10^{-9}}_{3.78*10^{-9}}$    & \vspace{6pt}  $3.95*10^{-24}\pm ^{1.38*10^{-22}}_{6.41*10^{-24}}$ \\

\hline \hline
\end{tabular}
\end{center}
\end{table}

\subsection{$f(R,\square R)$ gravity}
As an example we will consider the case where the Lagrangian
contains terms of the form $\Box R$. However, to keep the analysis
simple we will consider only the first order term of such
corrections and in this case the Lagrangian will be given by \be
f(R,R_{\mu \nu} R^{\mu \nu},\Box R)= R+ \xi \Box R \ee In this
case the extra term $\Box R$ can be rewritten as a total
divergence and, as expected, does not contribute at all in the
field equations. This can also be seen from the Friedmann Eq.
(\ref{friedeq1}), which in this special case simplifies to the
usual Friedmann equation of Einstein gravity.

The next most interesting case in this family of theories is the
Lagrangian \be f(R,R_{\mu \nu} R^{\mu \nu},\Box R)= R+ \xi (\Box
R)^2 \ee Then, $G_{eff}$ is given by \be G_{eff}(a)=\frac{1}{8
\pi} \frac{1+8\frac{k^6}{a^6} \xi }{1+6 \frac{k^6}{a^6} \xi
}\label{geffex2}\ee All other cases involving terms $(\Box R)^n$
with $n>2$ give complicated functions that also involve $\Box R$
and thus are difficult to calculate.

As in the previous example, it is possible to use Eq.
(\ref{geffex2}) to find analytically the best and the $1\sigma$
values of $\xi$ according to BBN \be \xi_{BBN}=-\frac{a_{BBN}^6
\left(-1+\frac{G_{BBN}}{G_0}\right)}{2 k^6 \left(-4+3
\frac{G_{BBN}}{G_0}\right)} \label{xi2}\ee In Fig. 1 (right) we
show the plot of $G_{eff}$, given by Eq. (\ref{geffex2}), for the
values of the parameter $\xi$, which correspond to the central and
$1\sigma$ values allowed by the BBN bounds for a value of $k=0.002
Mpc^{-1}$. However, since the $k$-mode is actually unknown and can
only be rather arbitrarily chosen, we have also plotted the value
of $\xi$ versus $k$ in Fig. 2 (right), and in Table 1 we show
$\xi$ for various values of the scale $k$. Finally, as in the
previous case we will also consider the $k$-mode corresponding to
the horizon size at the BBN, and the corresponding constraints are
shown in Table 1.

Using the local gravity constraints for this example and taking
into consideration eq. (\ref{geffex2}), we demand that
$\frac{k^6}{a^6} |\xi| \ll 1$, which gives the following
constraint: \be |\xi| \ll \lambda_k^6 \sim 10^{-68} Mpc^6 \ee
Again this is more robust than the ones found by using the BBN
constraint, the latter are not excluded as $\xi_{BBN}$ has a
larger $1\sigma$ error region, so the two constraints overlap with
each other.

Another very interesting case of this class of theories is to
consider terms of the form $R \Box R$, instead of just $\Box R$.
These terms correspond to the first order correction of a
Lagrangian of the form $R+\sum_{n=0}^\infty c_n R \Box^n R$, which
were shown in Ref. \cite{Biswas:2005qr} to give rise to a ghost
and asymptotically free theory of gravity. Thus, keeping only the
first order correction the Lagrangian is \be f(R,R_{\mu \nu}
R^{\mu \nu},\Box R)= R+ \xi R \Box R \ee and $G_{eff}$ is given by
\be G_{eff}(a)=\frac{1+\xi \Box R -8 \frac{k^4}{a^4} \xi }{8 \pi
(1+\xi \square R) \left(1+\xi {\square R}-6 \frac{k^4}{a^4} \xi
\right)} \label{geffex3}\ee As can be seen by Eq. (\ref{geffex3}),
$G_{eff}$ also depends $\Box R$ instead of just the scale factor
$a$ like in the previous cases. Unfortunately, we were unable to
find either an analytical solution, as the Friedmann Eq.
(\ref{friedeq1}) in this case is a very complex fourth order
differential equation, or a numerical one as we do not have enough
initial conditions. Thus, we were unable to provide a constraint
for $\xi$ using the BBN bounds or plot $G_{eff}$ as a function of
the scale factor $a$.

\section{Conclusions}
Our analysis covers modified gravity models with a generic class
of Lagrangian density with higher order and terms of the form
$\frac{1}{2} f(R,R_{\mu \nu} R^{\mu\nu},\Box R)$. Using the fact
that at long ranges the linear perturbation analysis
differentiates Einstein's gravity with respect to any
modification, through the time evolution of the gravitational
constant, we derived the matter density perturbation equation and
the effective gravitational ``constant" $G_{eff}$ for the action
(\ref{action}).

We also used the BBN bounds on the gravitational constant, which
are of the order of $10\%$, in order to test the difference
between Einstein and modified gravity theories. The reason why the
BBN bounds can be used to test modified gravity theories is that
the value of the gravitational constant determines the expansion
rate of the Universe and thus the relevant time scales for the
production of light elements (H, He and Li). This fact allowed us
to test several cases of our general Lagrangian and constrain
their parameters. Furthermore, we applied the BBN constraints to
study the $\Box R$ correction in the Einstein-Hilbert action.

However, the fact that the values we found for the parameter $\xi$
are actually larger than one would expect, it means that the
energy scale at which these correction terms, e.g.
$R_{\mu\nu}R^{\mu\nu}$, are introduced is quite low. For instance
one could write the corresponding term in the Lagrangian as
$\frac{1}{M^2} R_{\mu\nu}R^{\mu\nu}$, where $\frac{1}{M^2}$ is the
parameter $\xi$. Now, one would naively expect $M$ to be of the
order of Planck scale or even higher, but in our case the value of
$M$ is much smaller than that.

This can be explained by the fact that presently the BBN bounds
have quite a large error themselves, which means that the
constraints we derived are not very strong. This can be seen by
the fact that the error on the derived parameter $\xi$ is quite
large, and this fact even allows for a zero value of $\xi$. Also,
the primordial nucleosynthesis is quite a complex phenomenon and
while its essence can be captured by a single data point, it is
certain that a complete analysis, i.e. one that would also include
the integration of the background equations from deep in the
radiation era up to today and the use of the proper nuclear
reaction rates, would most certainly provide stringent
constraints.

We have also implemented local gravity constraints, following the
approach of Ref. \cite{Tsujikawa:2007gd}. As expected, the new
constraints are more robust than the ones found by using the BBN
constraint; however, the latter are not excluded as $\xi_{BBN}$
has a larger $1\sigma$ error region, so the two constraints
overlap with each other.

\section*{Acknowledgements}
The authors would like to thank T.~Sotiriou for useful discussions
and for pointing out a minor error on Sec. IV. The authors
acknowledge support by the Niels Bohr International Academy and by
the EU FP6 Marie Curie Research $\&$ Training Network
``UniverseNet" under Contract No. MRTN-CT-2006-035863. S.N. also
acknowledges support by the Danish Research Council under FNU
Grant No. 272-08-0285.

\end{document}